\documentclass[conference]{IEEEtran}

\usepackage[letterpaper, left=0.680in, right=0.680in, bottom=1.049in, top=0.71in]{geometry}

\usepackage{ifpdf}
\ifCLASSINFOpdf
\usepackage[pdftex]{graphicx}
\graphicspath{{./Figures/}}
\DeclareGraphicsExtensions{.pdf,.jpeg,.png}
\else
\usepackage[dvips]{graphicx}
\DeclareGraphicsExtensions{.pdf}
\usepackage[caption=false, font=footnotesize]{subfig}
\fi

\usepackage{tabularx, booktabs}
\usepackage{amssymb}
\usepackage{amsthm} % for bold Lemma
\usepackage[ruled,vlined,linesnumbered]{algorithm2e}
\usepackage{aligned-overset} % for align overset
\usepackage{array}
\usepackage{balance}
\usepackage{cite}
\usepackage{color}
\usepackage{epstopdf}
\usepackage{enumitem} % for leftmargin
\usepackage{mathtools} % for \splitfrac
\usepackage{makecell} % for thick hline
\usepackage{multicol} % for two equations side-by-side
\usepackage{multirow} % for multiple sub-rows of a cell of a table
\usepackage[caption=false,font=small]{subfig}
\usepackage{textcomp}
\usepackage{stfloats}
\usepackage{url}
\usepackage{verbatim}
\usepackage{graphicx}
\usepackage{url}
\renewcommand{\vec}[1]{{\mathbf{#1}}}

\newtheorem{remark}{Remark}

\begin{document}

\title{Generative and Explainable AI for High-Dimensional Channel Estimation}

\author{
    \IEEEauthorblockN{Nghia~Thinh~Nguyen~and~Tri~Nhu~Do}
    \IEEEauthorblockA{ Department of Electrical Engineering, Polytechnique Montr\'{e}al, Montr\'{e}al, Qu\'{e}bec, Canada. }
    \IEEEauthorblockA{ Emails: \{nghia-thinh.nguyen, tri-nhu.do\}@polymtl.ca }
	}
\maketitle

\begin{abstract}
In this paper, we propose a new adversarial training framework to address high-dimensional instantaneous channel estimation in wireless communications. Specifically, we train a generative adversarial network to predict a channel realization in the time-frequency-space domain, in which the generator exploits the third-order moment of the input in its loss function and applies a new reparameterization method for latent distribution learning to minimize the Wasserstein distance between the true and estimated channel distributions. Next, we propose an explainable artificial intelligence mechanism to examine how the critic discriminates the generated channel. We demonstrate that our proposed framework is superior to existing methods in terms of minimizing estimation errors. Additionally, we find that the critic's attention focuses on the high-power portion of the channel's time-frequency representation.
\end{abstract}


\vspace{-1em}
\section{Introduction}

Channel state information (CSI) acquisition, particularly channel estimation in 5G New Radio (NR), presents unprecedented challenges, especially in handling high-dimensional data \cite{Hirzallah2021TCCN}. The increased degrees of freedom (DoF) related to the dimension of transmission pilots across the time, frequency, and space domains lead to higher computational complexity with traditional methods \cite{Ma2021IEEE}. In this context, generative artificial intelligence (AI) emerges as a potential solution to these pressing challenges \cite{Ye2018WCL}. Recently, researchers have focused on enhancing realistic 5G NR pilot-based channel estimation techniques through advanced simulation frameworks like Sionna \cite{Sionna2022}, following standards such as 3GPP 38.901. However, the shift from conventional methods like Least Square (LS) and Linear Minimum Mean Square Error (LMMSE) to AI-driven approaches in high-dimensional scenarios introduces its own set of research challenges \cite{Jiao2024IEEE}.

Generative adversarial network (GAN)-based channel estimation has been proposed as an alternative solution. \cite{Haider2024TVT} employs a GAN-based approach as the primary method to estimate the channel for high degrees of freedom in terms of channel components. \cite{Du2023MDPI} uses a GAN-based method combined with pilot information to estimate the channel. \cite{Balevi2021TWC} utilized a Wasserstein GAN with gradient penalty (WGAN-GP) as a component to estimate the channel based on the LMMSE method. However, a fundamental issue with the GAN model is the imbalance between the critic and the generator, as the critic converges too quickly, leading the generator to produce a limited variety of samples or experience vanishing gradients. This results in the generator failing to learn effectively from the critic's feedback, manifesting as inaccurate or incomplete reconstruction of channel state information \cite{Bau2019ICCV}.

\textit{In this paper}, we propose a novel framework for adversarial training aimed at instantaneous channel estimation. Specifically, we focus on enhancing the generator's capability by leveraging additional probabilistic characteristics of the input features. To achieve this, we introduce an innovative loss function that integrates the third-order moment of the channel impulse response (CIR) with reconstruction loss, thereby capturing higher-order statistics of the channel characteristics. Additionally, we aim to strengthen the latent distribution learning capability of semi-supervised methods, such as Variational Autoencoder (VAE). Based on the proposed framework, we pose a \textit{research question}: How can the AI model, particularly the critic in our framework, interpret the channel gain across the time, frequency, and space domains to effectively discriminate the generated channel?

\textit{The contributions} of this work are twofold\footnote{The source code developed for this paper is available at \url{tnd-lab.work}}. ($i$) We propose the use of the third-order moment of high DoF input in the generator’s loss function and introduce a novel method for VAE's latent distribution learning, resulting in superior performance, as demonstrated by lower normalized mean square error (NMSE) in channel estimation. Notably, our estimation framework does not require pilot knowledge and, in some cases, achieves better performance than traditional methods. ($ii$) We introduce a new explainable AI method using activation mapping for the critic, termed AM4C. We observe a significant alignment between the critic's attention and the high-power gain regions in the time-frequency representation (TFR) of the CIR, providing insights into the design of the generator’s loss, pilot patterns, and other loss functions for future research.

\vspace{-1em}
\section{System Model}

\noindent{\textbf{Notations.}} 
\textit{Underlined notations} denote random variables, such as a single scalar $\underline{\alpha}$, a vector $\vec{\mathbf{\underline{x}}}$, or a matrix $\vec{\mathbf{\underline{X}}}$, while \textit{non-underlined notations} represent realizations of these random variables, e.g., $\alpha$, $\vec{x}$, and $\vec{X}$.

\subsubsection{System Components and Network Topology}

We consider a random network topology to analyze our proposed channel estimation framework comprehensively. Uplink (UL) estimation is performed by a single-antenna user terminal (UT) communicating with a base station (BS) equipped with $K_{\rm ant}$ antennas.
The coordinates of the UT and BS in spherical coordinates are denoted as $\vec{u}_{\rm ut,bs} = [d_{\rm ut,bs}, \varphi_{\rm ut,bs}, \phi_{\rm ut,bs}]$, where $d_{\rm ut,bs}$ is the radial distance, $\varphi_{\rm ut,bs}$ is the elevation angle of arrival along the $z$-axis, and $\phi_{\rm ut,bs}$ is the azimuth angle in the horizontal $x$-$y$ plane. The UT antenna follows an omnidirectional pattern, so the antenna response vector for the UT is represented as $a_{\rm ut}(\varphi, \phi) = 1$.
The BS is equipped with a uniform linear array (ULA) of $K_{\rm ant}$ cross-polarized antennas. The UL array response vector is:
$\vec{a}_{\rm bs}(\varphi, \phi) = [a_{1}(\varphi, \phi) e^{\psi_{1}}, \ldots, a_{K_{\rm ant}}(\varphi, \phi)e^{\psi_{K_{\rm ant}}}]$, 
where $a_{n}(\varphi, \phi)$ represents the individual gain and phase response of the $n$-th antenna, $\lambda_{cf}$ is the wavelength of the carrier frequency, and $\psi_{n} = - j \frac{2\pi}{\lambda_{cf}} (n-1) d_y \sin(\varphi) \sin(\phi)$ denotes the phase shift of the $n$-th antenna, $d_y$ is the distance between two antennas.

\begin{figure}[t]
    \centering
    \includegraphics[width=\linewidth]{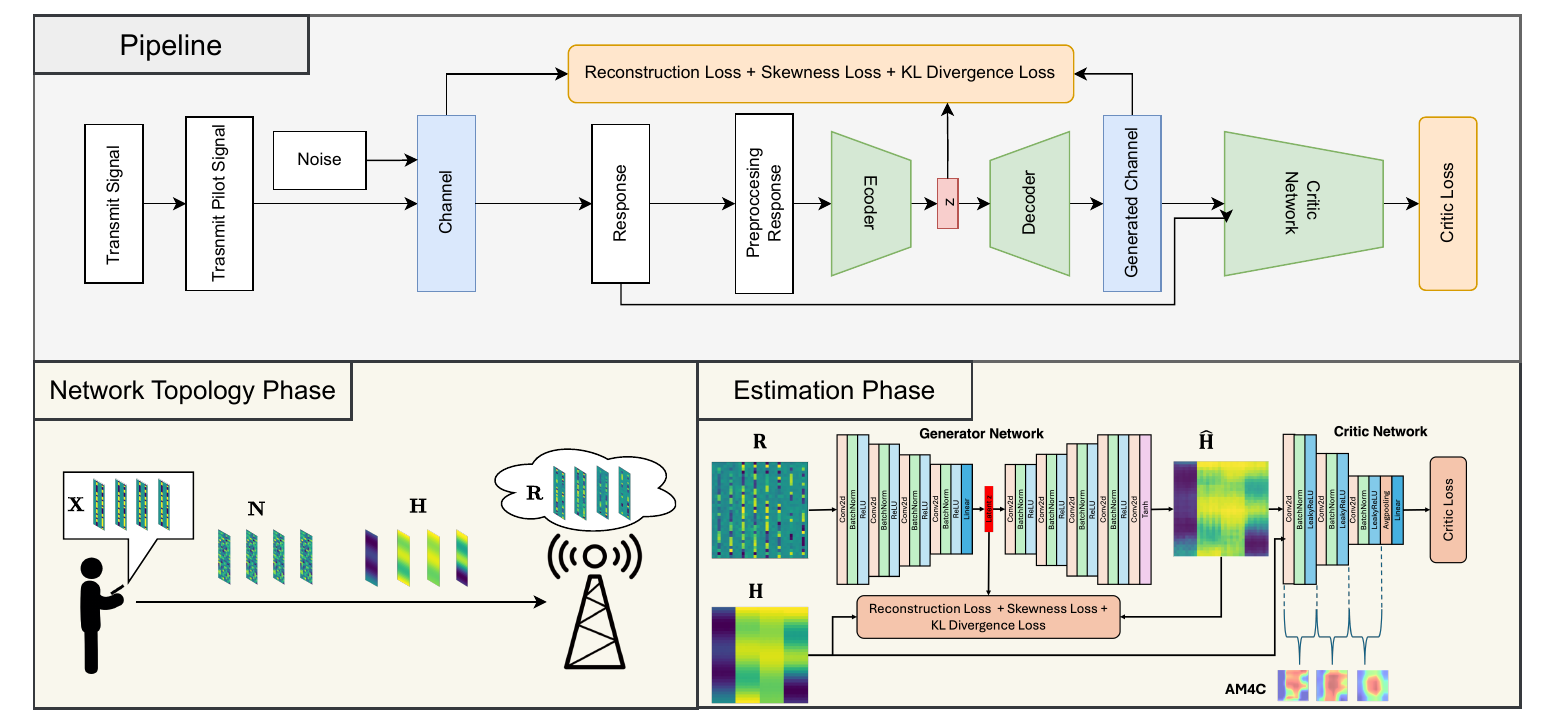}
    \caption{Proposed WGAN-GP framework with explainable AI mechanism.}
    \label{fig:pipleline}
    \vspace{-1em}
\end{figure}

\subsubsection{Signal Modeling}

% The user transmits a signal with the binary random bit $ b_{intput} \in \{0,1\}^{K_{bit}} $ for each symbol transmitter, where $ K_{bit} $ is the total number of bit. The Mapper function $\mathcal{M}(\cdot)$ translates bits into complex symbols according to the QAM constellation \cite{Sionna2022}, which is represented as $ x_{mapper} = \mathcal{M}(b_{input}) $. Consider the number of subcarriers $ K_{\rm sc} $, the number of OFDM Symbols $ K_{\rm sym} $, and two-bit per symbol. The number of transmitted bits for the signal antenna of the UT can be computed as $ K_{bit} = K_{\rm sc} \times K_{\rm sym}  $. The assumption is that after two time steps of total $ K_{\rm sym} $ time step, i.e time step $ \{2, 6\} $ with $ K_{\rm sym} = 8 $, we choose the pilots symbol $ \vec{p} $ follow up the Kronecker pilot pattern  with $ \vec{p} \in \mathbb{C}^{2 \times 1} $. 
Let $ K_{\rm{sym}} $ be the number of OFDM symbols, i.e., time domain, $ K_{\rm{sc}} $ be the umber of effective subcarriers, i.e., frequency domain. Recall that $ K_{\rm{ant}} $ is the number of antennas, representing the space domain. 
A pilot mask, $\mathbf{M} \in \{0,1\}^{K_{\rm sc} \times K_{\rm sym}}$, is created using the orthogonal pilot pattern with Kronecker structure \cite{Sionna2022}. Let $ \mathcal{T} \subseteq \{0, 1, \dots, N_{\rm{sym}}-1\} $ be the set of OFDM symbol indices reserved for pilot transmission, $ \mathcal{K}_{i,j} \subseteq \{0, 1, \dots, N_{\rm{sc}}-1\} $ be the set of subcarrier indices allocated to transmitter $ i $ and stream $ j $ for pilot transmission.
Considering \textit{sparse} pilot pattern, the mask exhibits sparseness into distinct time and frequency components, i.e., $\mathbf{M} = \mathbf{P}_t \otimes \mathbf{P}_f^{i,j}$, where $ \mathbf{P}_t(s) $ is a binary vector indicating pilot OFDM symbol indices, $ \mathbf{P}_f^{i,j}(k) $ is a binary vector indicating the subcarrier indices allocated to transmitter $ i $ and stream $ j $ \cite{Sionna2022}.
Here, $\vec{X}_p \in \mathbb{C}^{K_{\rm sc}\times K_{\rm sym}}$ can represented as
$\mathbf{X}_p^{i,j}(k,s) = p_{i,j}$ if $\mathbf{M}^{i,j}(k,s) = 1$, and $0$ otherwise, where $ p_{i,j} $ are the QAM pilot symbols for transmitter $ i $, stream $ j $.
$\mathbf{M}$ is designed as $\mathbf{M}^{i,j}(k,s) = 1$ if $s \in \mathcal{T}, k \equiv i K_{\text{str}} + j \ (\text{mod}\ K_{\text{seq}})$, and $0$ otherwise. Combining the mask and symbol assignment, the complete pilot pattern $ \mathbf{X} \in \mathbb{C}^{K_{\rm sc}\times K_{\rm sym}}$ within the resource grid (RG) can be expressed as $\mathbf{X}^{i,j}(k,s) = \mathbf{M}^{i,j}(k,s) \times \mathbf{X}_p^{i,j}(k,s)$.

\subsubsection{Channel Modeling}

We consider communication between the UT and BS using the Urban Micro (UMi) channel model \cite{Sionna2022}. Let \( h_{n, s, k} \) denote the channel for the \( n \)-th antenna and the \( k \)-th subcarrier at time step \( s \). With the applied network topology, it can be represented as
\begin{align} \label{eq_channel_1ant}
    &h_{n, k,s} = \textstyle\sum_{\ell=0}^{K_{\rm path}-1} h_{\ell}(s) e^{-j2\pi k \Delta_f \tau_{\ell}} a_{n}(\varphi_{\ell}, \phi_{\ell}) e^{\psi_{n,\ell}}
    ,
\end{align}
where \( K_{\rm path} \) denotes the number of multi-path components, and \( \Delta_f \) represents the subcarrier spacing. Here, \( h_{\ell}(s) \sim \mathcal{CN}(0,1) \) denotes the complex amplitude, \( \tau_{\ell} \) the delay, and \( \psi_{n,\ell} \) the phase shift, all provided by the UMi model \cite{Sionna2022}.
According to \eqref{eq_channel_1ant}, the instantaneous TFR of the channel \( \vec{H} \in \mathbb{C}^{K_{\rm ant} \times K_{\rm sym} \times K_{\rm sc}} \) represents a single realization of \( \vec{\mathbf{\underline{H}}} \). 

% Let me know if you need any further adjustments!

\subsubsection{Received Observation Modeling}

Based on the channel model and signal modeling, assuming that the cyclic prefix (CP) is added to the transmit waveform and removed from the received waveform in the time domain, the TFR of the received signal, \( \vec{\mathbf{\underline{R}}} \), at the BS for the UT is expressed as
\begin{align}
    \vec{\mathbf{\underline{R}}} = \vec{\mathbf{\underline{H}}} \odot \vec{\mathbf{\underline{X}}}_{\text{exd}} + \vec{\mathbf{\underline{N}}},
\end{align}
where \( \vec{\mathbf{\underline{R}}} \in \mathbb{C}^{K_{\rm ant} \times K_{\rm sym} \times K_{\rm sc}} \) represents the received signal, and \( \vec{\mathbf{\underline{N}}} \in \mathbb{C}^{K_{\rm ant} \times K_{\rm sym} \times K_{\rm sc}} \) is an i.i.d. matrix of additive white Gaussian noise (AWGN) with entries \( [\vec{\mathbf{\underline{N}}}]_{ij} \sim \mathcal{CN}(0,1) \).
The expanded version of \( \vec{\mathbf{\underline{X}}} \) is denoted by \( \vec{\mathbf{\underline{X}}}_{\text{exd}} = \text{repmat}(\vec{\mathbf{\underline{X}}}, K_{\rm{ant}}, 1, 1) \), where \( \text{repmat} \) replicates \( \vec{\mathbf{\underline{X}}} \) along the antenna dimension.

\section{Proposed Adversarial Training Framework}

\subsection{Adversarial Training for Channel Estimation}

As aforementioned, we address the channel estimation problem using an adversarial training framework consisting of a generator $G$ that generates synthetic channel estimates and a critic $D$ that discriminates between the generated channel and the true channel. The generator $ G $ receives the observation $ \vec{\underline{R}} $ as input, mapping it as $ G (\vec{\underline{R}}) $ such that
$
    G (\vec{\underline{R}}): \mathbb{R}^{K_{\rm ant} \times K_{\rm sym} \times K_{\rm sc} } \rightarrow \mathbb{R}^{K_{\rm ant} \times K_{\rm sym} \times K_{\rm sc} } 
$,
where $\vec{\underline{R}} \mapsto  \widehat{\vec{\underline{H}}}$.
Thus, the instantaneous channel estimate is $\widehat{\vec{H}} = G (\vec{R})$. 
Given the true random instantaneous channel $ \vec{\underline{H}} $, the input of the critic can be either $ \vec{\underline{H}} $ or $ \widehat{\vec{\underline{H}}} $. The critic can be represented as
$
    D(\vec{H}^{*}) : \mathbb{R}^{K_{\rm{ant}} \times K_{\rm{sym}} \times K_{\rm{sc}} } \rightarrow [0,1]
$
where $\vec{\underline{H}}^{*} \mapsto \{ p(\mathbf{H}^*|H_0), p(\mathbf{H}^*|H_1) \}$. 
Here, $ \vec{\underline{H}}^{*} $ can represent either $ \vec{\underline{H}} $ or $ \widehat{\vec{\underline{H}}} $ channel model. The likelihood probability $ p(\mathbf{H}^*|H_0) $ and $ p(\mathbf{H}^*|H_1) $ corresponding to the hypothesis $ H_0 $ for the generated channel and $ H_1 $ for the true channel.

The overall objective of the GAN framework is a \textit{two-player minimax game}. The critic tries to maximize the value function $  V(D, G)  $ to distinguish between true and generated channels, while the generator tries to minimize this same value function to deceive the critic. 
Let $ p_{\vec{\mathbf{\underline{H}}}}(\vec{H}) $,  $ p_{\widehat{\vec{\mathbf{\underline{H}}}}}(\widehat{\vec{H}}) $, $ p_{\vec{\mathbf{\underline{R}}}}(\vec{R}) $ denote the distribution of the true channel, the generated channel, and the received observation, respectively. The minimax game of our channel estimation problem is expressed as
\begin{align} 
    &\min_G \max_D V(D, G) = \nonumber\\
    &\mathbb{E}_{\vec{H} \sim p_{\vec{\mathbf{\underline{H}}}}(\vec{H})} [ \log D(\vec{H})] \!\!+\!\! \mathbb{E}_{\vec{R} \sim p_{\vec{\mathbf{\underline{R}}}}(\vec{R})} [ \log(1 \!\!-\!\! D(G(\vec{R})))] \nonumber \\
    &= \mathbb{E}_{\vec{H} \sim p_{\vec{\mathbf{\underline{H}}}}(\vec{H})} [ \log D(\vec{H})] \!\!+\!\! \mathbb{E}_{\widehat{\vec{H}} \sim p_{\mathbf{\widehat{\vec{\mathbf{\underline{H}}}}}}(\widehat{\vec{H}})} [ \log(1 \!\!-\!\! D(\widehat{\vec{H}}))]
    .
\end{align}
The technical challenge is the imbalanced optimization between the critic and generator networks during the early training phase. The critic $D$ output has a more direct objective, discriminating tasks constrained to $ [0, 1] $, while the generator $G$ learns to produce channel distributions over a wider and more complex domain, i.e., $[\min \vec{\mathbf{\underline{H}}}, \max \vec{\mathbf{\underline{H}}}]$. In addition, $\vec{\underline{R}}$ and $\vec{\underline{H}}$ are high-dimensional data, which makes it challenging for $G$ to learn the complex underlying distribution.

\vspace{-.5em}
\subsection{Problem Formulation for Optimizing the Generator}

The training objective in our paper is to enable the generator to converge quickly from one distribution to another, i.e., from $p_{\widehat{\vec{\mathbf{\underline{H}}}}}(\widehat{\vec{H}})$ to $p_{\vec{\mathbf{\underline{H}}}}(\vec{H})$. To achieve this, the training process aims to minimize their Wasserstein distance, i.e., $ W (p_{\vec{\mathbf{\underline{H}}}}(\vec{H}), p_{\widehat{\vec{\mathbf{\underline{H}}}}}(\widehat{\vec{H}})) $.
Since the set of all joint distributions $ \prod (p_{\vec{\mathbf{\underline{H}}}}(\vec{H}), p_{\widehat{\vec{\mathbf{\underline{H}}}}}(\widehat{\vec{H}})) $ is not available, Kantorovich-Rubinstein's dual formulation is used to simplify the Wasserstein distance \cite{Gulrajani2017WGANGP}.
\begin{align}
    &W(p_{\vec{\mathbf{\underline{H}}}}(\vec{H}), p_{\widehat{\vec{\mathbf{\underline{H}}}}}(\widehat{\vec{H}})) \notag \\
    &=  \sup_{\|f\|_\mathcal{L}\leq 1} \big\{ \mathbb{E}_{ \vec{{H}}\sim p_{\vec{\mathbf{\underline{H}}}}(\vec{H})}[f( \vec{H} )] - \mathbb{E}_{ \widehat{\vec{{H}}} \sim p_{{\widehat{\vec{\mathbf{\underline{H}}}}}} (\widehat{\vec{H}})}[f( \widehat{\vec{H}})]
    \big\}
    ,
\end{align}
where $f(\cdot)$ is a function parameterized by a neural network. Even if the probability density functions $ p_{\widehat{\vec{\mathbf{\underline{H}}}}} (\widehat{\vec{H}}) $ and $ p_{\vec{\mathbf{\underline{H}}}}(\vec{H}) $ do not perfectly align, the Wasserstein distance between them can remain small as long as their supports are close and the mass (probability density) is similarly distributed within those supports. In other words, if the generated channel estimates $ \widehat{\vec{H}} $ have a distribution whose support largely overlaps with that of the true channel $ \vec{H} $ and the mass is distributed in a similar fashion, then the Wasserstein distance will be small. Therefore, our goal is to ensure that the distribution $ p_{\widehat{\vec{\mathbf{\underline{H}}}}} (\widehat{\vec{H}}) $ of the estimated channels closely matches the distribution $ p_{\vec{\mathbf{\underline{H}}}}(\vec{H}) $ of the true channels, both in terms of support and mass distribution. To this end, first we let $ \widehat{\vec{H}} $ and $ \vec{H} $ be normalized to the same scale, i.e., the same range, such as $ \vec{H}$ and $ \widehat{\vec{H}}$ being in $ [-1, 1] $. Considering pilot sparsity and adversarial training, the proposed channel estimation problem is formulated as
\begin{subequations}\label{minimizes_channel_estimation}
\begin{alignat}{2}
G^* \!=\! &\quad \arg \min_{(G | D)}
& % empty column
% & \mathbf{E} \left\{ \|\vec{H} - \widehat{\vec{H}})\|^2 \right\} \\
& W(p_{\vec{\mathbf{\underline{H}}}}(\vec{H}), p_{\widehat{\vec{\mathbf{\underline{H}}}}}(\hat{\vec{H}})) \\
&\text{subject to}    
% & & K_{\rm sym} \leq K_{\rm sym}^{\max},\\
% & & & [\vec{p}]_i \in \mathcal{C}, \forall i \\
& &  \mathfrak{Re}\{[\vec{H}]_{ij}\}, \mathfrak{Im}\{[\Vec{H}]_{ij}\}, \mathfrak{Re}\{[\widehat{\vec{H}}]_{ij}\}, \nonumber\\
&&& \mathfrak{Im}\{[\widehat{\vec{H}}]_{ij}\}  
% \sim \mathcal{N}(\mu_{ij}, \sigma^2_{ij})
\in [-1,1], \forall i, \forall j , \label{constraint_scaled_data}\\
% &&& \lim_{\text{epoch} \to \text{max-iteration}}\Pr(\widehat{\vec{H}}|\vec{H}) \to 1 \\
% &&& \lim_{\text{epoch} \to \text{max-iteration}}\Pr(\widehat{\vec{H}}|\vec{H}) \to 0 
% \label{minimizes_channel_estimation1} \\
% &&& |\mathcal{T}| \leq K_{\rm sym} \label{constraint_pilots}\\
% &&& |i-j|> 1 \\
&&& | p_{i,j} |^2  \leq 1, i \leq |\mathcal{T}|, j \leq K_{\rm sc}, 
\label{constraint_pilots}
\end{alignat}
\end{subequations}
where $(G|D)$ indicates that we focus on optimizing $G$ given a specific $D$. Constraint \eqref{constraint_scaled_data} represents data normalization, while constraint \eqref{constraint_pilots} enforces sparsity in the pilot pattern. We set $p_{i,j} = \pm \frac{1}{\sqrt{2}} \pm \mathsf{j} \frac{1}{\sqrt{2}}$, ensuring that $|p_{i,j}|^2 = 1$.
Note that while the realization values are scaled to $[-1, 1]$, the distribution of the true channel $ p_{\vec{\mathbf{\underline{H}}}}(\vec{H}) $ and the generated channel $ p_{\widehat{\vec{\mathbf{\underline{H}}}}}(\widehat{\vec{H}}) $ differ. Achieving alignment between these distributions remains challenging.

\section{Third-Order Moment-Based VAE-WGAN-GP}

\subsection{Mathematical Description of the Proposed G and D }

\subsubsection{NN Architecture} 

Our goal is to upgrade the generator to a more refined version. Based on the work of \cite{Gulrajani2017WGANGP}, further investigated in \cite{Arjovsky2017TPMTGAN}, we define the generator function $ f_{\rm{g}} $ with the input being the observation data $ \vec{\underline{R}} $, and the set of parameters $ \boldsymbol{\theta} $ is updated accordingly. Note that for a composite function, $f\circ g = f(g(x))$, and $\vec{H}^{*} = f_{\rm{g}}(\vec{R}; \boldsymbol{\theta})$.
Let $f_{\rm{g}}^{[bl_{i}]}$ denote the $i$-th \textit{feature extraction block}, which can be represented as $f_{\rm{g}}^{[bl_{i}]} = f_{\rm{g}}^{[a_i]} \circ f_{\rm{g}}^{[b_i]} \circ f_{\rm{g}}^{[c_i]}$, comprising activation, batch normalization, and convolution layers. Drawing from the VAE architecture \cite{Kingma2013CoRR}, the $G$ function can be divided into the encoder, latent space, and decoder parts.
Let $\vec{a}_{\rm{g,enc}} = f_{\rm{g,enc}}(\vec{R}; \boldsymbol{\theta}_{\vec{a}_{\rm{g,enc}}}) = f_{\rm{g,enc}}^{[bl_{1}]} \circ f_{\rm{g,enc}}^{[bl_{2}]} \circ f_{\rm{g,enc}}^{[bl_{3}]} \circ f_{\rm{g,enc}}^{[bl_{4}]} \circ f_{\rm{g,enc}}^{[l]}$ represent the output vector of the encoder function. The latent vector $\vec{z}$ is the output of the transformation function $\vec{z} = f_{\rm{g,trans}}(\vec{a}_{\rm{g,enc}})$ and serves as the input to the decoder function. The decoder output is given by $\widehat{\vec{H}} = f_{\rm{g,dec}}(\vec{z}; \boldsymbol{\theta}_{\rm{g, dec}}) = f_{\rm{g,dec}}^{[bl_{1}]} \circ f_{\rm{g,dec}}^{[bl_{2}]} \circ f_{\rm{g,dec}}^{[bl_{3}]} \circ f_{\rm{g,dec}}^{[bl_{4}]} \circ f_{\rm{g,dec}}^{[\tanh]}$. The function $f_{\rm{g,dec}}^{[\tanh]}$ ensures that the output is normalized within $[-1, 1]$, aligning with the scaling of the true channel $\vec{H}$.
Thus, $G$ is mathematically described as
\begin{align} \label{eq_math_generator}
    &f_{\rm{g}}(\vec{R}; \mathcal{\boldsymbol{\theta}}) = f_{\rm{g,dec}}(f_{\rm{g,trans}}(f_{\rm{g,enc}}(\vec{R}; \boldsymbol{\theta}_{\vec{a}_{\rm{g,enc}}})); \boldsymbol{\theta}_{\rm{g, dec}})
    .
\end{align}

In contrast, the critic network, denoted as $ f_{\rm{c}} $, has a set of parameters $ \boldsymbol{\omega} $ and includes a feature extraction block defined as $f_{\rm{c,ft}}^{[bl_{i}]} = f_{\rm{c}}^{[a_i]} \circ f_{\rm{c}}^{[b_i]} \circ f_{\rm{c}}^{[c_i]}$ at the $i$-th block. Thus, the critic $D$ can be mathematically expressed as
\begin{align} \label{eq_math_critic}
    &f_{\rm{c}}(\vec{H}^{*}; \boldsymbol{\omega}) = f_{\rm{c}}^{[l_4]}(f_{\rm{c}}^{[avg_4]}(f_{\rm{c,ft}}^{[bl_3]}(f_{\rm{c,ft}}^{[bl_2]}(f_{\rm{c,ft}}^{[bl_1]}(\vec{H}^{*};\boldsymbol{\omega}_{\rm{c, ft}}^{[bl_1]}); \nonumber\\
    &\qquad\qquad\qquad\qquad \boldsymbol{\omega}_{\rm{c, ft}}^{[bl_2]});\boldsymbol{\omega}_{\rm{c, ft}}^{[bl_3]});\boldsymbol{\omega}_{\rm{c}}^{[avg_4]}), \boldsymbol{\omega}_{\rm{c}}^{[l_4]})
    .
\end{align}

Before training the critic network, the function $ f_{\rm{c}}^{[\rm dim]}(\vec{H}^{}): \mathbb{R}^{K_{\rm ant} \times K_{\rm sym} \times K_{\rm sc} } \mapsto \mathbb{R}^{(K_{\rm ant} \times K_{\rm sym}) \times K_{\rm sc} } $ is defined to reduce the dimensionality of the channel input $ \vec{H}^{} $, making the model easier to train and reducing computational costs for high-dimensional channels. As the critic’s target, the output of the linear function $ f_{\rm{c}}^{l_4} $ is scalar, with $ f_{\rm{c}}^{l_4}(X_{\vec{\underline{H}}}^{[l_4]}, \boldsymbol{\omega}_{\rm{c}}^{[l_4]}) > f_{\rm{c}}^{l_4}(X_{\widehat{\vec{\underline{H}}}}^{[l_4]}, \boldsymbol{\omega}_{\rm{c}}^{[l_4]}) $, where $ X_{\vec{\underline{H}}}^{[l_4]} $ and $ X_{\widehat{\vec{\underline{H}}}}^{[l_4]} $ are the inputs to the linear function for $ D(\vec{H}) $ and $ D(\widehat{\vec{H}}) $, respectively.

\subsubsection{Proposed Latent Distribution Learning in VAE's Probabilistic Framework} $q(\vec{z}|\vec{H})$ represent the true posterior probability given the real channel $\vec{H}$. Since $\vec{H}$ cannot be observed directly, $q(\vec{z}|\vec{H})$ cannot be computed directly. We approximate it by sampling the approximate posterior probability of the encoder output vector, where $q(\vec{z}|\vec{H}) \approx p(\vec{z}|\vec{a}_{\rm{g,enc}})$. Using the reparameterization method, the latent vector $\vec{z}$ is defined as $\vec{z} = f_{\rm{g,trans}}(\vec{a}_{\rm{g,enc}}) = \mu_{\vec{a}_{\rm{g,enc}}} + \epsilon \sigma_{\vec{a}_{\rm{g,enc}}}$, where $\vec{z} \sim p(\vec{z}|\vec{a}_{\rm{g,enc}}) = \mathcal{N}(\mu_{\vec{z}, \mathcal{\boldsymbol{\theta}}}, \sigma_{\vec{z}, \mathcal{\boldsymbol{\theta}}})$. In most cases, the estimated latent vector $\vec{z} \sim \mathcal{N}(0,1)$ is used for the inference phase. However, in our scenario, $\widehat{\vec{H}} \sim q(\widehat{\vec{H}}|z)$ cannot be generated randomly by the decoder using a Gaussian random vector as input.
Moreover, $\vec{a}_{\rm{g,enc}}$ is a compressed vector of the observation $\vec{R}$ rather than $\vec{H}$, making $q(\vec{z}|\vec{H}) \approx p(\vec{z}|\vec{a}_{\rm{g,enc}})$ unreasonable, as $p_{\vec{\underline{R}}}(\vec{R})$ and $p_{\vec{\underline{H}}}(\vec{H})$ differ, as evidenced by the results. The proposed objective is to optimize the latent vector $\vec{z} \sim p(\vec{z}|\vec{a}_{\rm{g,enc}})$ to closely approximate $p(\vec{z}|\vec{H})$, with $p_{\vec{\underline{H}}}(\vec{H}) \sim \mathcal{N}(\mu_{\vec{\underline{H}}}, \sigma_{\vec{\underline{H}}}^2)$.
It is assumed that 
the number of training sets of real channel $|\mathcal{D}_{\text{train}}|$ is large enough to compute $\mu_{\vec{\underline{H}}}$, and $\sigma_{\vec{\underline{H}}}$. The latent vector $\vec{z} \sim \mathcal{N}(\mu_{\vec{z}, \mathcal{\boldsymbol{\theta}}}, \sigma_{\vec{z}, \mathcal{\boldsymbol{\theta}}}^2)$ is estimated directly from compressed vector $\vec{a}_{\rm{g, enc}}$ and it is transformed close to $\mathcal{N}(\mu_{\vec{\underline{H}}}, \sigma_{\vec{\underline{H}}}^2)$. Applying trick parameterization, $\vec{z} = f_{\rm{g,trans}}(\vec{a}_{\rm{g,enc}}) = \mu_{\vec{\underline{H}}} + \vec{a}_{\rm{g,enc}} \sigma_{\vec{\underline{H}}}$. 

\subsection{Proposed Loss Function}

The target critic function $ D $  tries to maximize the distance of two distributions, while the generator model $ G $ is the opposite.

\subsubsection{Loss of D} The loss function of critic network $ D $ is minimized by updating $\boldsymbol{\omega}$. Its formulation is represented as 
\begin{align} 
    L_{D} &= \underbrace{\mathbb{E}_{\widehat{\vec{H}}\sim p_{\widehat{\vec{\mathbf{\underline{H}}}}}(\widehat{\vec{H}})}[D_{\boldsymbol{\omega}}(\widehat{\vec{H}})]}_{\triangleq L_{\tt f} {\text{: generative loss}}}
    - \underbrace{\mathbb{E}_{\vec{H} \sim p_{\vec{\mathbf{\underline{H}}}}(\vec{H})}[D_{\boldsymbol{\omega}}(\vec{H})]}_{\triangleq L_{\rm r} {\text{: true loss}}}  
    \label{eq_L_D_a}
    .
\end{align}

Consider $ p_{\tilde{\vec{\mathbf{\underline{H}}}}}(\tilde{\vec{H}}) $ as the mixture distribution of the $ p_{\vec{\mathbf{\underline{H}}}}(\vec{H}) $ and $ p_{\widehat{\vec{\mathbf{\underline{H}}}}}(\widehat{\vec{H}}) $, is given by:
$
    p_{\tilde{\vec{\mathbf{\underline{H}}}}}(\tilde{\vec{H}}) = \epsilon p_{\vec{\mathbf{\underline{H}}}}(\vec{H}) + (1 - \epsilon) p_{\widehat{\vec{\mathbf{\underline{H}}}}}(\widehat{\vec{H}})
$
where $  \epsilon \sim \mathcal{U}(0, 1) $ is a random scalar, drawn from a uniform distribution between 0 and 1. To constrain the critic network's update speed and preserve the 1-Lipschitz property. The gradient penalty loss can be defined as
\begin{align}\label{eq_L_gp}
    L_{\rm gp} = \lambda_{\rm gp} \mathbb{E}_{\tilde{\vec{H}}\sim p_{\tilde{\vec{\mathbf{\underline{H}}}}}(\tilde{\vec{H}})}[(\|\nabla{D_{\boldsymbol{\omega}}(\tilde{\vec{H}})}\|_2 - 1)^2] 
    .
\end{align}
From \eqref{eq_L_D_a} and \eqref{eq_L_gp}, the adopted $L_D$ is expressed as
\begin{align}
    L_{D} &= \text{generative loss} - \text{true loss} + \text{gradient penalty} \nonumber\\
    &= L_{\rm f} - L_{\rm r} + L_{\rm gp} 
    .
    \label{eq_L_D_e}
\end{align}

\subsubsection{Proposed Loss of G} 

In the minimax game framework, the discriminator \( D \) function seeks to maximize the Lipschitz continuity of its output by controlling the gradients of \( \widehat{\vec{H}} \) through its parameters \( \boldsymbol{\omega} \). Concurrently, the generator \( G \) updates its parameters \( \boldsymbol{\theta} \) to minimize the discrepancy between the generated data and real data. Initially, the loss of $G$ can be designed based on the negative fake loss \cite{Gulrajani2017WGANGP} as
\begin{align}
    L_{G} = - L_{\rm f} = - \mathbb{E}_{\vec{R} \sim p_{\vec{\mathbf{\underline{R}}}}(\vec{R})}[D(G_{\boldsymbol{\theta}}(\vec{R}))]
    \label{eq_L_G_a}
    .
\end{align}
Taking into account the evidence lower bound of the VAE generator, we consider the KL divergence loss and the reconstruction loss, respectively, as
\begin{align}
    L_{\rm KL} &\!=\! \lambda_{\rm KL}\frac{1}{2}\!\! \sum_{d=1}^{\vec{z}_{\rm dim}} \!\! \left( \log \frac{\sigma_{H_d}^2}{\sigma_{z_d, \boldsymbol{\theta}}^2} \!\!+\!\! \frac{\sigma_{z_d, \boldsymbol{\theta}}^2 \!\!+\!\! (\mu_{z_d, \boldsymbol{\theta}} \!\!-\!\! \mu_{H_d})^2}{\sigma_{H_d}^2} \!\!-\!\! 1 \! \right),
    \label{eq_kl}
    \\
L_{\rm rec} &= \mathbb{E}_{\vec{H} \sim p_{\vec{\mathbf{\underline{H}}}}(\vec{H}), \vec{R} \sim p_{\vec{\mathbf{\underline{R}}}}(\vec{R})}[\|\vec{H} - G_{\boldsymbol{\theta}}(\vec{R})\|^2]
    \label{eq_rec_loss}
    .
\end{align}
Next, we aim to make the two distributions $ p_{\vec{\mathbf{\underline{H}}}}(\vec{H}) $ and $ p_{\tilde{\vec{\mathbf{\underline{H}}}}}(\tilde{\vec{H}}) $ as close as possible. Let $ \gamma_{\rm{r}} $ and $ \gamma_{\rm{g}} $ represent the \textit{coefficients of skewness} or the third-order moments of the true and estimated channel distributions, respectively, which can be expressed as
\begin{align}
    \gamma_{\rm{r}} &= \left(\mathbb{E}_{\vec{H} \sim p_{\vec{\mathbf{\underline{H}}}}(\vec{H})}[(\vec{H} - \mu_{\rm{r}})^3]\right) / \sigma_{\rm{r}}^3 
    \label{eq_gamma_r}
    ,
    \\
    \gamma_{\rm{g}} &= \left(\mathbb{E}_{\widehat{\vec{H}} \sim p_{\widehat{\vec{\mathbf{\underline{H}}}}}(\widehat{\vec{H}})}[(\widehat{\vec{H}} - \mu_{\rm{g}})^3]\right) / {\sigma_{\rm{g}}^3}
    \label{eq_gamma_g}
    ,
\end{align}
where $ \mu_{\rm{r}} $, $ \mu_{\rm{g}} $, $ \sigma_{\rm{r}} $, and $ \sigma_{\rm{g}} $ are the mean and the standard deviation of the true and generative distribution respectively. It is assumed that the distance of the skewness of true and generative distributions is close and continuous if the distance of two distributions is close. Thus, from \eqref{eq_gamma_r} and \eqref{eq_gamma_g}, we can define our third order moment-based loss as 
\vspace{-.5em}
\begin{align}
    L_{\gamma} &= |\gamma_{\rm{r}} - \gamma_{\rm{g}}|
    \label{eq_skewness_loss} 
    .
\end{align}
\vspace{-.5em}
From \eqref{eq_L_G_a}-\eqref{eq_rec_loss}, and \eqref{eq_skewness_loss}, the proposed $L_G$ is expressed as
\begin{align}
    L_{G} &= - \text{ fake loss} + \text{rec loss} + \text{kl loss} +  \text{skewness loss} \nonumber\\
    &= - L_{\rm f} + L_{\rm rec} + L_{\rm KL} + L_{\gamma}
    .
    \label{eq_G_loss_final}
\end{align}
The proposed VAE-WGAN-GP-based channel estimation method, with enhanced $G$ loss and latent distribution learning, is outlined in the pseudo-code of Algorithm 1.

\begin{algorithm}[htbp]
\SetAlgoLined
\KwIn{Noisy received signal $ \vec{R} \sim p_{\vec{\mathbf{\underline{R}}}}(\vec{R}) $. The true channel data $ \vec{H} \sim p_{\vec{\mathbf{\underline{H}}}}(\vec{H}) $ }
\KwOut{Trained Generator $  G  $ for channel estimation $ \widehat{\vec{H}} \sim p_{\widehat{\vec{\mathbf{\underline{H}}}}}(\widehat{\vec{H}}) $ }
Compute $\mu_{\vec{\underline{H}}}$, $\sigma_{\vec{\underline{H}}}$ from training set $\mathcal{D}_{\text{train}}$\;
\For{$e = 0, \dots, K_{epoch}$}{
    Sample $ \{\vec{H}_{i}^{[e]}\}^{b}_{i=1}\ \sim p_{\vec{\mathbf{\underline{H}}}}(\vec{H}) $ a batch from true channel\;
    Sample $ \{\vec{R}_{i}^{[e]}\}^{b}_{i=1}\ \sim p_{\vec{\mathbf{\underline{R}}}}(\vec{R}) $ a batch from received signal\;
    \For{$ t = 0, \dots, n_{c}$}{
        $ \{\widehat{\vec{H}}_{i}^{[t], [e]}\}^{b}_{i=1} \leftarrow G(\{\vec{R}_{i}^{[e]}\}^{b}_{i=1}) \sim p_{\widehat{\vec{\mathbf{\underline{H}}}}}(\widehat{\vec{H}})$ predict a batch from generated channel\;
        $ \{\tilde{\vec{H}}_{i}^{[t], [e]}\}^{b}_{i=1}\ \leftarrow \epsilon \{\vec{H}_{i}^{[e]}\}^{b}_{i=1} + (1 - \epsilon) \{\widehat{\vec{H}}_{i}^{[t], [e]}\}^{b}_{i=1}$ a batch of mixture channel\;
        Training $a_{\tt {c}}^{[t], [e]} \leftarrow D(\{\vec{H^{*}}_{i}^{[t], [e]}\}^{b}_{i=1})$\;
        Compute $f_{L_{D}}(a_{\tt {c}}^{[t], [e]})$\;
        Compute $f_{L_{gp}}(\{\tilde{\vec{H}}_{i}^{[t], [e]}\}^{b}_{i=1})$\;
        $f_{L_{D}}^{[t], [e]} \leftarrow f_{L_{D}}(a_{\tt {c}}^{[t], [e]}) + f_{L_{gp}}(\{\tilde{\vec{H}}_{i}^{[t], [e]}\}^{b}_{i=1})$\;
        $ \boldsymbol{\omega}^{[t+1], [e]} \leftarrow \text{Opt}(f_{L_{D}}^{[t], [e]}, \boldsymbol{\omega}^{[t], [e]}, \eta, \beta_1, \beta_2) $\; 
    }
    $\vec{a}_{\rm{g, enc}}^{[e]} \leftarrow f_{\tt {g, enc}}(\{\vec{R}_{i}^{[e]}\}^{b}_{i=1}) \sim p(\vec{a}_{\rm{g, enc}})$\;
    Reparameterize $\vec{z}^{[e]} \leftarrow \mu_{\vec{\underline{H}}} + \vec{a}_{\rm{g, enc}}^{[e]}\sigma_{\vec{\underline{H}}} \sim p(\vec{z}|\vec{a}_{\rm{g, enc}})$\;
    $\{\widehat{\vec{H}}_{i}^{[e]}\}^{b}_{i=1} \leftarrow f_{\tt {g, dec}}(\vec{z}^{[e]})$\;
    Compute $f_{L_{f}}(\{\widehat{\vec{H}}_{i}^{[e]}\}^{b}_{i=1})$\;
    Compute $f_{L_{rec}}(\{\widehat{\vec{H}}_{i}^{[e]}\}^{b}_{i=1}, \{\vec{H}_{i}^{[e]}\}^{b}_{i=1})$\;
    Compute $f_{L_{\rm KL}}(\vec{z}^{[e]}, \mu_{\vec{\underline{H}}}, \sigma_{\vec{\underline{H}}})$\;
    $f_{L_{G}}^{[e]} \leftarrow f_{L_{f}}(\{\widehat{\vec{H}}_{i}^{[e]}\}^{b}_{i=1}) + f_{L_{rec}}(\{\widehat{\vec{H}}_{i}^{[e]}\}^{b}_{i=1}, \{\vec{H}_{i}^{[e]}\}^{b}_{i=1}) + f_{L_{\rm KL}}(\vec{z}^{[e]}, \mu_{\vec{\underline{H}}}, \sigma_{\vec{\underline{H}}})$\;
    $ \boldsymbol{\theta}^{[e+1]} \leftarrow \text{Opt}(f_{L_{G}}^{[e]}, \boldsymbol{\theta}^{[e]}, \eta, \beta_1, \beta_2) $\;
}
\textbf{Initialization:} Initialize parameters of $  G  $ and $  D  $, set learning rates $ \eta $, the batch size $ b $,  the number of iterations of the critic per generator iteration $ n_{\rm{c}} $, other Optimizer's parameters $ \beta_1 $ and $ \beta_2 $ 

\caption{
Training Procedure of the Proposed VAE-WGAN-GP Generator and Critic Models}
\end{algorithm}

\begin{remark}
    Numerical results demonstrate that all components of the Generator loss $ L_G $, including the fake loss $ L_{\rm f} $, reconstruction loss $ L_{\rm rec} $, KL divergence loss $ L_{\rm KL} $, and the third-order moment-based skewness loss $ L_\gamma $, successfully converge during the training process. This collective convergence of the individual loss components ensures the overall convergence of the proposed Generator loss $ L_G $, as illustrated in Fig.~\ref{fig:fig_loss}.
\end{remark}

\subsection{Inference and Performance Analysis}

To infer the model, our strategy involves simulating $|\mathcal{D}_{\text{test}}|$ samples from the Sionna framework, with each sample containing instantaneous channels with added noise as test sets. To adapt the model architecture, we reshape the actual channel based on $ f_{\rm{c}}^{[\rm dim]}(\vec{H}^{*}) $. 
The evaluation metric is the Normalized Mean Square Error (NMSE), which measures the difference between the true and generated channels, defined as
\begin{align}
    \text{NMSE} = \frac{\sum_{i=1}^{|\mathcal{D_{\text{test}}}|} (\vec{H}_i - \widehat{\vec{H}}_i)^2}{\sum_{i=1}^{|\mathcal{D_{\text{test}}}|} (\vec{H}_i)^2}
    .
\end{align}

\subsection{Explainable AI: Activation Mapping  for the Critic (AM4C)}

We propose an Explainable AI method to study the interpretability of the critic $D$ during the training loop. Specifically, utilizing the activation mapping method, we propose the AM4C algorithm to identify and visualize the regions in the TRF of the channel that significantly influence the decision of $D$.
The activation map of the $k$ feature map at the spatial location $(i, j)$ from the convolution function $f_{\rm{c}}^{[c_n]}$, where $n \in \{1, 2, 3\}$ is expressed as 
\begin{align}
    f_{\rm{c}}^{[c_n]}(\mathbf{X}_{\rm{c}}^{[c_n]}; \boldsymbol{\omega}_{\rm{c}}^{[c_n]}) = \{\vec{A}_{k,i,j}^{c_n} \mid k \in \{1, \ldots, K\}\}\}
\end{align}
where $ K $ is the number of spatial locations, $\mathbf{X}_{\rm{c}}^{[c_n]}$ is input of convolution layer. 
Next, we compute the gradient of the critic output of $ f_{\rm{c}}(\vec{H}^{*}; \boldsymbol{\omega} $ with respect to each feature map’s activation $\frac{\partial{f_{\rm{c}}(\vec{H}^{*}; \boldsymbol{\omega}}}{\partial{f_{\rm{c}}^{[c_n]}(\mathbf{X}_{\rm{c}}^{[c_n]}; \boldsymbol{\omega}_{\rm{c}}^{[c_n]})}}$.  Let $ a^{k}_{\rm{c}} $ be the weight representing the overall influence of the critic decision value as follows
\begin{align}
    a^{k}_{\rm{c}} = \textstyle \frac{1}{K} \sum_{i, j} \frac{\partial{f_{\rm{c}}(\vec{H}^{*}; \boldsymbol{\omega}})}{\partial{f_{\rm{c}}^{[c_n]}(\mathbf{X}_{\rm{c}}^{[c_n]}; \boldsymbol{\omega}_{\rm{c}}^{[c_n]})}}
    .
\end{align}
Consider $ \mathtt{ReLU} $ activation function to remove negative values. The activation map $ M_{D}^{c_{n}}{(i, j)} $ can be defines as
\begin{align}
    M_{D}^{c_{n}}{(i, j)}
    = \textstyle \mathtt{ReLU} 
    &\big( (1/K) \sum_{k} \sum_{i, j} \frac{\partial{f_{\rm{c}}(\vec{H}^{*}; \boldsymbol{\omega}})}{\partial{f_{\rm{c}}^{[c_n]}(\mathbf{X}_{\rm{c}}^{[c_n]}; \boldsymbol{\omega}_{\rm{c}}^{[c_n]})}} \nonumber \\
    &\quad\times f_{\rm{c}}^{[c_n]}(\mathbf{X}_{\rm{c}}^{[c_n]}; \boldsymbol{\omega}_{\rm{c}}^{[c_n]}\big)
    .
\end{align}

The resulting activation map $ M_{D}^{c_{n}}(i, j) $ provides a visual and quantitative representation of which spatial regions in the input channel data are most influential in the critic $D$. 
High values in the activation map indicate areas with substantial impact, thereby serving as the \textit{attention regions} that the model focuses on during channel estimation.

\begin{remark}
    After training, the attention areas identified by the critic $D$ consistently align with regions of high magnitude in the channel's TRF, as demonstrated in Fig.~\ref{fig_cm}.
\end{remark}
\section{Numerical Results and Visualization}
Key simulation settings are presented in Table I; additional settings for simulation, training, and inference can be found in our published source code.
\vspace{-1em}
\begin{table}[h]
\centering
\caption{Network, Simulation, and Hyper-parameters Settings}
\begin{tabular}{ll || ll}
\hline
 Parameter & Value & Parameter & Value \\ \hline
$K_{\rm ant}$ & 4 & $\lambda_{\rm gp}$ & 10\\ 
$K_{\rm sym}$ & 8  & $\lambda_{\rm KL}$ & 0.1\\
$K_{\rm sc}$ & 32 & $\mathcal{T}$ & $\{2,6\}$ \\
% $K_{\rm epoch}$ & 100  & $\eta$ & $0.002$ \\ 
 % $|\mathcal{D}_{\text{test}}|$ & 2000 & $b$ & 256 \\ 
 % $|\mathcal{D}_{\text{train}}|$ & 16000 & $n_{\rm{c}}$ & 3 \\
\hline
\end{tabular}
\end{table}

\subsubsection{Training Evaluation}
\begin{figure}[htp!]
    \centering
    \includegraphics[width=.7\linewidth]{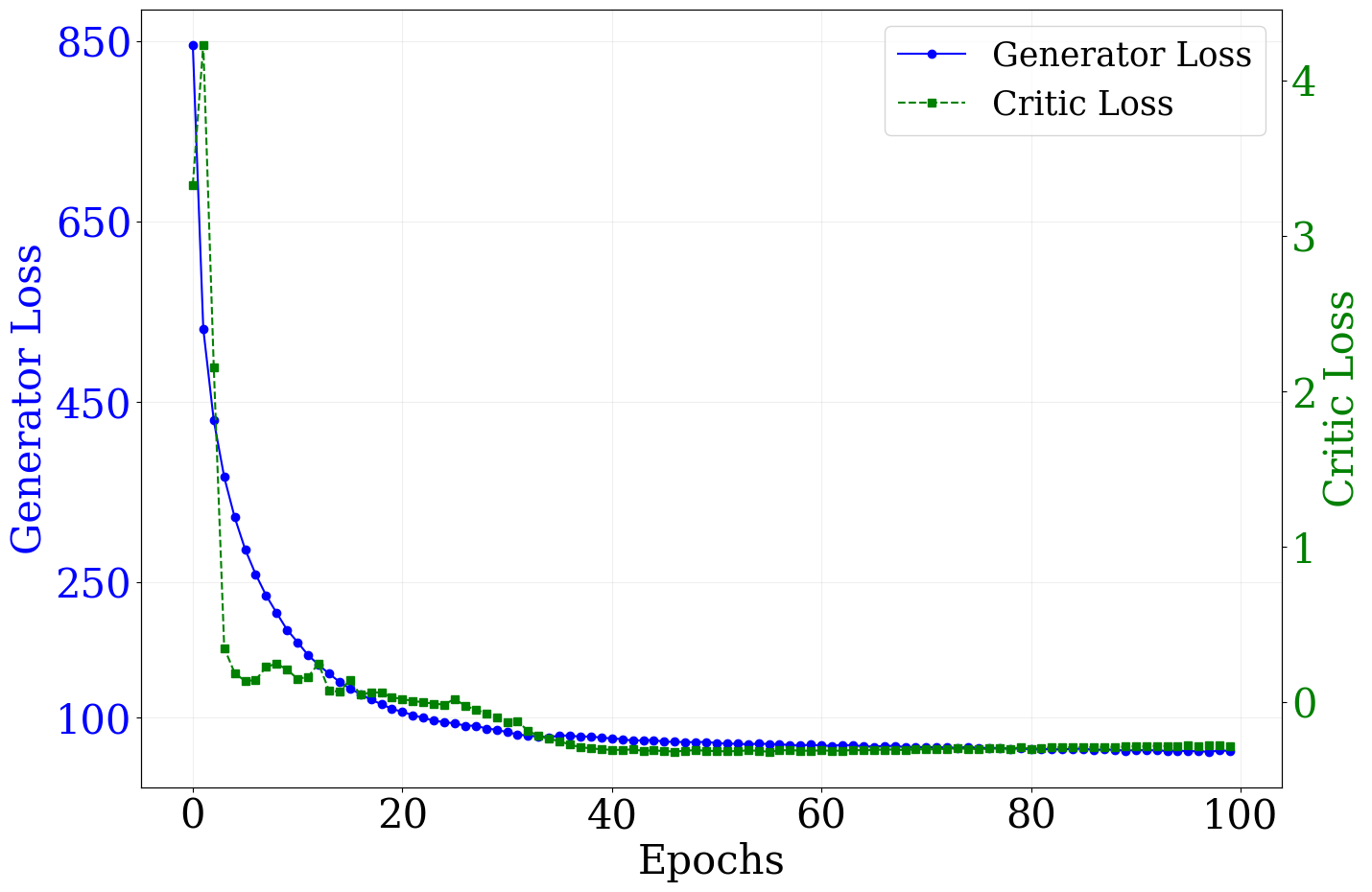}
    \caption{Loss function of $G$ and $D$}
    \label{fig:fig_loss}
\end{figure}

To demonstrate the advantage of our proposed method, Fig.~\ref{fig:fig_loss} shows that the generator loss and the citic loss tend to decrease shapely after $K_{epoch}$. It means that the proposed loss function convergence. 
% To demonstrate the advantage of our proposed method, Fig.~\ref{fig_g_losses} shows that the generator loss of the WGAN-GP-based model tends to slightly decrease because it is not dependent on $ p_{\vec{\mathbf{\underline{H}}}}(\vec{H}) $. This means that it cannot reconstruct $ p_{\widehat{\vec{\mathbf{\underline{H}}}}}(\widehat{\vec{H}}) \sim p_{\vec{\mathbf{\underline{H}}}}(\vec{H}) $. In contrast, when the generator loss includes reconstruction loss and skewness loss, it tends to approach zero. Additionally, with the skewness loss, the generator achieves better performance.

% Similarly, in Fig.~\ref{fig_c_losses}, the critic loss function fluctuates near zero after a few epochs because the critic is being fooled by the generator. However, when a conditional loss is applied, the critic becomes much more stable and converges very quickly.

% \begin{figure}[t]
% 	\centering
% 	\subfloat[]{%
% 		\includegraphics[width=.7\linewidth]{Figures/WGAN_G_Loss_2.png}
% 		\label{fig_g_losses}} \hfill
% 	\subfloat[]{%
% 		\includegraphics[width=.7\linewidth]{Figures/WGAN_C_Loss_2.png}
% 		\label{fig_c_losses}} 
% 	\caption{Convergence of (a) Generator's losses and (b) Critic's losses.}
% 	\label{fig_cm} 
% \end{figure}   

\subsubsection{Inference}

\begin{figure}[htp!]
    \centering
    \includegraphics[width=.8\linewidth]{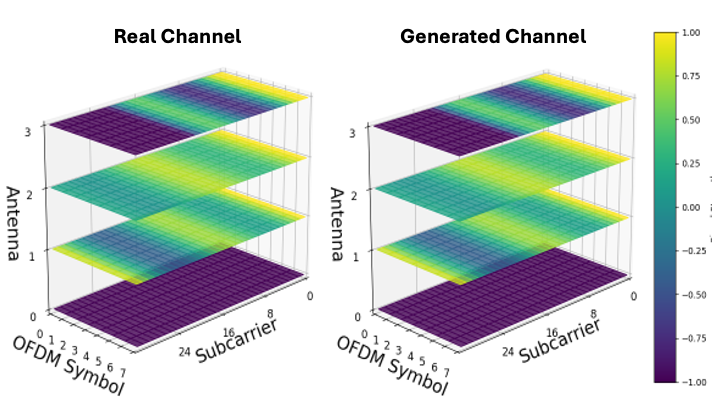}
    \caption{True 3D channel realization and corresponding generative channel.}
    \label{fig:normalied_channel_result}
\end{figure}

For testing, we use the same topology as the training set. To ensure fidelity, we also generate instantaneous channels with random noise as a test set and then evaluate the dataset. The comparison is shown in Fig.~\ref{fig:normalied_channel_result}; the magnitude of the channel on each antenna is reconstructed similarly to the true channel. The results also indicate that the normalization process impacts the reconstruction of channel characteristics.

\subsubsection{Performance Evaluation in terms of NMSE}

\begin{figure}[htp!]
    \centering
    \includegraphics[width=\linewidth]
    {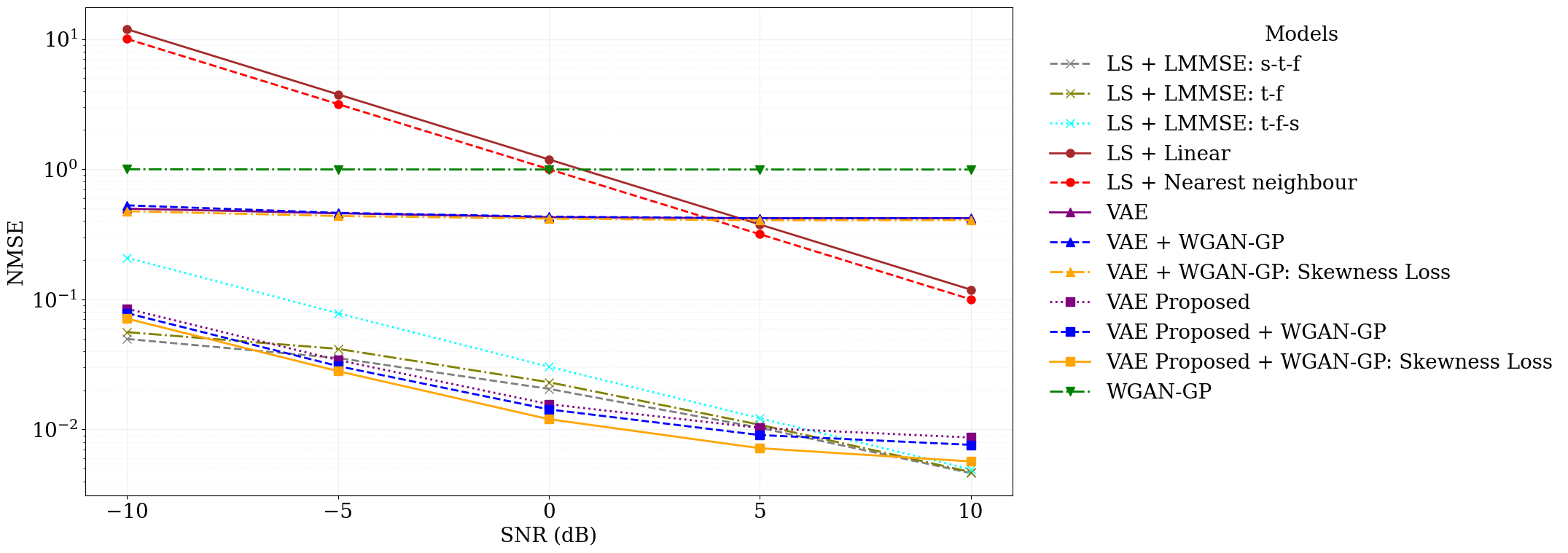}
    \caption{NMSE of the proposed VAE-WGAN-GP and conventional methods.}
    \label{fig:fig_nmse_score}
\end{figure}

We compare the NMSE of the proposed method with conventional methods using LS with interpolation techniques, i.e., nearest neighbor, linear, and LMMSE across different data domains. Fig.~\ref{fig:fig_nmse_score} shows that our methods obtain comparable outcomes with LMMSE methods in the lower SNRs. Besides it is a note that learning to generate samples without explicit distribution of WGAN-GP gives a poor result if compared with VAE methods. However, combining VAE and WGAN-GP can be a better solution, even with the proposed loss function which is the best result in our experiment.

% which use WGAN-GP with additional losses, achieve the best performance. In particular, with the skewness loss, the NMSE reaches the lowest score of $4.5 \times 10^{-5}$ at $0$ dB, while conventional techniques such as LMMSE estimation achieve an NMSE of $8 \times 10^{-3}$ at the same SNR level. From $0$ to $25$ dB, our model's NMSE tends to increase slightly because we trained the model by overfitting the dataset at $0$ dB SNR. This occurs because the model is trained on a fixed $0$ dB, causing overfitting on other datasets with higher SNRs.

% \begin{figure}[htp!]
%     \centering
%     \includegraphics[width=\linewidth]{Figures/fig4_v6.pdf}
%     % {Figures/nmse_snr_4ant_updated.png}
%     \caption{NMSE of the proposed WGAN-GP and conventional methods.}
%     \label{fig:fig_nmse_score}
% \end{figure}

\begin{figure}[htp!]
    \centering
    \includegraphics[width=\linewidth]
    {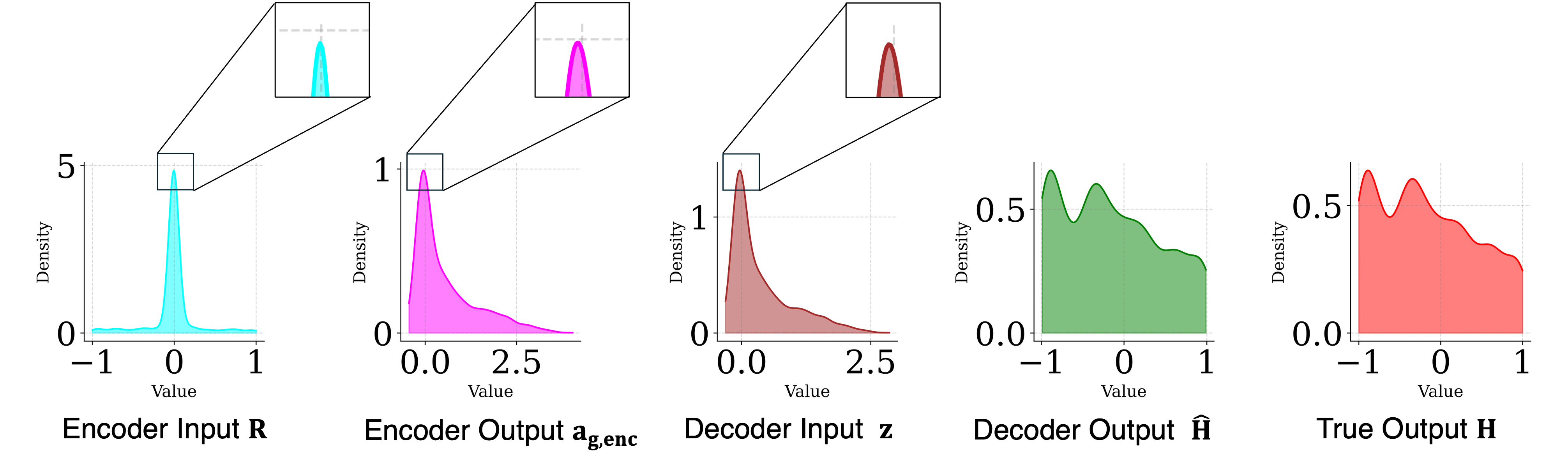}
    \caption{The distribution of input and output of the model, $p_{\vec{\underline{R}}}(\mathbf{R})$,  $p(\vec{a}_{\rm{g, enc}})$, approximated posterior $p(\vec{z}|\vec{a}_{\rm{g, enc}})$, $p_{\vec{\underline{\widehat{H}}}}(\mathbf{\widehat{H}})$, $p_{\vec{\underline{H}}}(\mathbf{H})$ }
    \label{fig:fig_distribution}
    \vspace{-1em}
\end{figure}

It is true that the encoder input $\vec{\underline{R}}$ estimate from the function of Gaussian noise which is slightly shifted by the operation of $\vec{\underline{X}}$ containing mostly zero. It means that $\vec{\underline{R}} \sim \mathcal{N}(\mu_{\vec{\underline{R}}},\sigma_{\vec{\underline{R}}})$, where $\mu_{\vec{\underline{R}}} \approx 0$, and $\sigma_{\vec{\underline{R}}} \approx 1 $ as can be seen from the first image of Fig.~\ref{fig:fig_distribution}. The distribution of the encoder output $\vec{a}_{\rm{g, enc}}$ is estimated as the same shape as the decoder input $\vec{z}$ and they are a right-skewed distribution. As a consequence, The distribution of the generated $\widehat{\vec{\underline{H}}}$ and the true $\vec{\underline{H}}$ is similar.

% \begin{figure}[htp!]
%     \centering
%     \includegraphics[width=\linewidth]
%     {Figures/fig_mu_sigma.png}
%     \caption{The distribution of mu and std of channel on training set}
%     \label{fig:fig_mu_sigma}
% \end{figure}

\subsubsection{Explainable AI for Attention Area of the Channel Matrix}

\begin{figure}[t]
	\centering
	\subfloat[]{%
		\includegraphics[width=.9\linewidth]{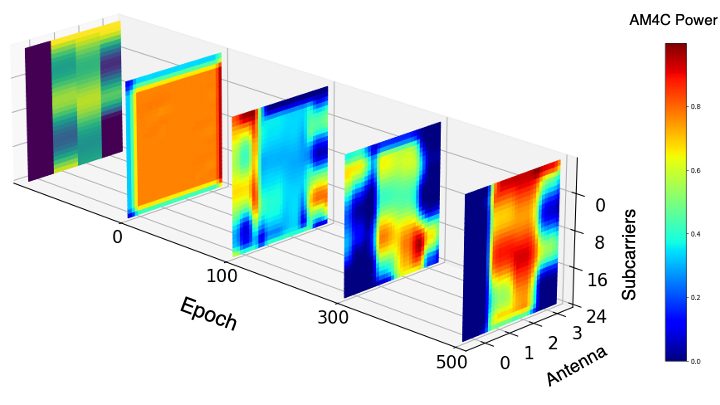}
		\label{fig:fig_am4c_conv_layer_1}} \hfill
	\subfloat[]{%
		\includegraphics[width=.9\linewidth]{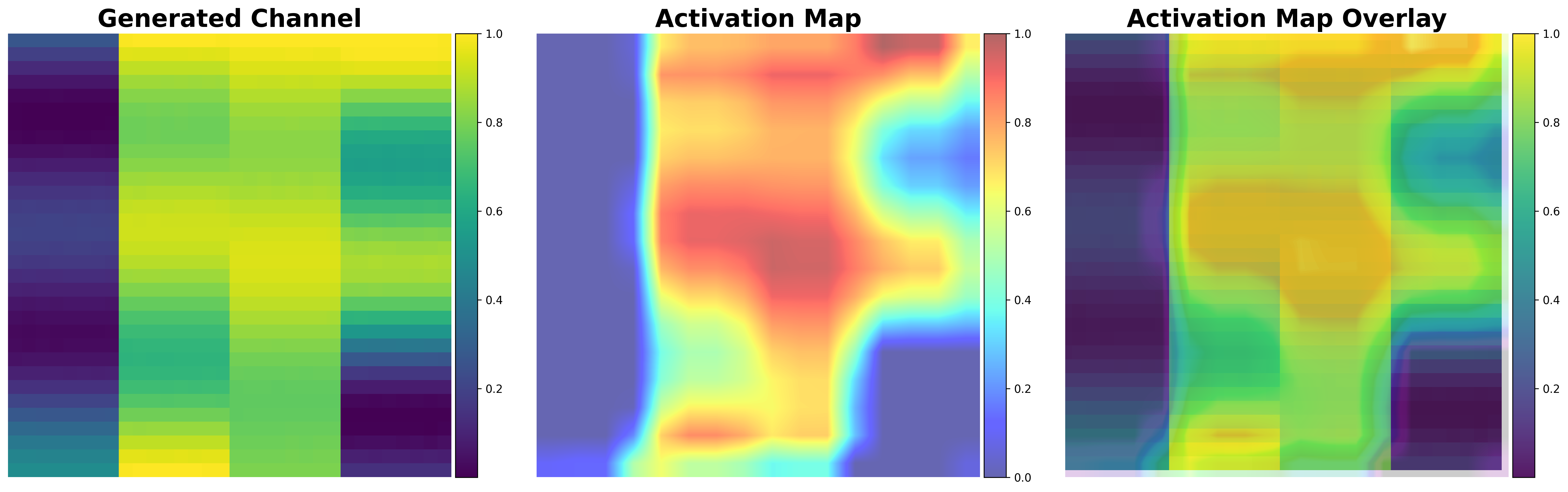}
		\label{fig:fig_am4c_conv_layer_2}} 
	\caption{(a) The evolution of the attention area of the critic over training epochs. (b) Attention aera of the trained critic overlayed over generative channel.}
	\label{fig_cm} 
 \vspace{-1em}
\end{figure}

In Fig.~\ref{fig:fig_am4c_conv_layer_1}, we illustrate how the generator model updates features during training and how the critic model utilizes these features to determine whether they are generative. At epoch zero, it is observed that the critic focuses on the entire channel due to the initial noise distribution. By epoch $100$, the critic tends to focus on lower magnitudes, as the generator easily produces these values and the critic seeks to avoid being misled by the generator. However, from epoch $300$ to $500$, the generator nearly surpasses the critic, prompting the critic to shift its focus to higher magnitudes, which contain more valuable information for classification.
The activation map, in Fig.~\ref{fig:fig_am4c_conv_layer_2}, originates from the first convolutional layer of the critic, indicating that this layer is the closest to extracting information from the generator. This suggests that the critic model focuses on higher magnitudes of resource elements in the TFR to for decision-making.

\vspace{-.5em}
\section{Conclusions}

In this paper, we propose a new WGAN-GP model for channel estimation, in which the generator exploits the third-order moment of the data to improve its performance. We also propose AM4C as an explainable AI mechanism to obtain the attention area of the critic. We show that the proposed WGAN-GP performs superiorly, achieving lower NMSE than conventional methods, particularly in favorable training settings. Using the proposed AM4C, we demonstrate that the critic relies on the high power gain regions of the time-frequency representation (TFR) of the channel to learn and discriminate the generated channel from the true channel. 

\bibliographystyle{IEEEtran}
\bibliography{References}

\end{document}